\documentclass[journal]{IEEEtran}
\usepackage{graphicx}

\usepackage{caption}

\usepackage{subcaption}

\usepackage{cite}

\usepackage[utf8]{inputenc}

\ifCLASSINFOpdf
\else
\fi
\usepackage{amsmath}
\usepackage{algorithmic}
\usepackage{algorithm}

\usepackage{bm}

\usepackage{array}

  \usepackage[caption=false,font=normalsize,labelfont=sf,textfont=sf]{subfig}
\usepackage{fixltx2e}
\begin{document}
%
\title{Secrecy Enhancement for UAV-enabled Integrated Sensing and Communication System}
%
%
%

\author{Chaedam~Son,~\IEEEmembership{Student Member,~IEEE,}
        Seongah Jeong,~\IEEEmembership{Member,~IEEE}
\thanks{This work was supported by the National Research Foundation of Korea space (NRF) grant funded by the Korea government (MSIT) (No. 2023R1A2C2005507)} 
\thanks{Chaedam Son and Seongah Jeong are with the School of Electronics Engineering, Kyungpook National University, Daegu 14566, Korea (e-mail: scd18347@gmail.com, seongah@knu.ac.kr).}
}

\maketitle

\begin{abstract}
In this correspondence, we propose an unmanned aerial vehicle (UAV)-enabled integrated sensing and communication (ISAC) system, where a full-duplex UAV equipped with uniform planar array (UPA) is adopted as a base station for the multiuser downlink communications, while sensing and jamming a passive ground eavesdropper. The goal of this work is to maximize the sum secrecy rate of ground users subject to the constraints of sensing accuracy and UAV's operational capability by jointly optimizing the transceiver beamforming and UAV's trajectory. 
To this end, we develop the algorithmic solution based on block coordinate descent (BCD) and semidefinite programming (SDP) relaxation techniques, whose performance is verified via simulations indicating its efficacy in improving communication security with the sufficient mission period. 
\end{abstract}


%
\IEEEpeerreviewmaketitle

\section{Introduction}
\IEEEPARstart{U}{nmanned} aerial vehicles (UAVs) have attracted  significant attention in cellular communications to support the explosive growth of data traffic and to improve the quality of the network services thanks to their flexible mobility and cost effectiveness. In UAV communications with the dominant line-of-sight (LOS) paths, it is challenging to maintain the privacy. To address the security issue, the physical-layer security (PLS) techniques have been actively investigated for UAV communications \cite{four}\cite{sensing}.

The evolution of 5G toward beyond 5G (B5G) and 6G leads a new potential development trend, that is, integrated sensing and communication (ISAC) systems owing to the  use of millimeter wave (mmWave) and terahertz (THz) that are preferred for the  high-resolution sensing applications
\cite{one}. Inspired by this trend, there are several literatures \cite{ten}\cite{seven} that investigate the PLS techniques in UAV-enabled ISAC systems, where the free-movable UAVs allows the additional degree of freedom (DoF) in security and sensing performance. In \cite{ten}, the authors propose the UAV-enabled wireless networks to maximize the number of securely-served downlink users subject to tracking performance constraints and quality-of-service (QoS) requirements, for which the online resource allocation design is developed based on a Kalman filter to sense and predict the eavesdropping UAV.
Similarly, the real-time UAV trajectory design of secure ISAC systems is proposed in \cite{ten} for supporting flexible secure communications performance based on extended Kalman filtering.
Since the UAV-enabled ISAC systems are at the beginning stage, the recent related works mostly assume a static UAV\cite{ten} or single-antenna \cite{seven}, whose performances can be further enhanced by increasing the DoF with UAV's mobility or multiple antennas. \\
\indent Motivated by these observations, we propose a multiuser downlink ISAC system, where a full-duplex UAV with uniform planar array (UPA) is adopted as a base station to serve the multiple ground users and to detect a single passive eavesdropper for jamming. In this correspondence, we aim at maximizing the sum secrecy rate of ground users by jointly optimizing the transceiver beamforming and UAV's trajectory subject to sensing accuracy and UAV's operational capability requirements. To this end, we develop the algorithmic solution based on block coordinate descent (BCD) method and semidefinite (SDP) relaxation\cite{convexopt},\cite{sixteen}. Notably, our work is distinct as it considers the full-duplex UAV-enabled secure ISAC systems that provides the valuable insights into the future UAV communications or ISAC systems, not explored in existing literature.\\
\indent \textit{Notations}: In this paper, vectors and matrix are denoted boldface, respectively. $\mathbf{C}^{M\times 1}$ represents a complex vector of dimension $M$. For a scalar $x$, its absolute value is denoted by $|x|$, and $[x]^+$ means $\text{max}(x, 0)$. $\mathcal{CN} (\mu, \sigma^2)$ represents the complex Gaussian normal distribution with mean $\mu$ and variance $\sigma^2$. $\text{Tr}(\cdot)$, $\| \cdot \|$, $(\cdot)^{\text{T}}$, $(\cdot)^{\text{H}}$, and $(\cdot)^{-1}$ represent trace, Euclidean norm, transpose conjugate transpose, and inverse. $\lambda_{\mathrm{max}}(\cdot)$ is the largest eigenvalue of matrix, and $\pmb{v}_{\mathrm{max}}(\cdot)$ is its corresponding eigenvector. $\otimes$ represents Kronecker product and $\pmb{I}_{N}$ is the $N\times N$ identity matrix. 
\section{Signal and System Model}
\begin{figure}[t]
    \centering
    \includegraphics[width=7cm]{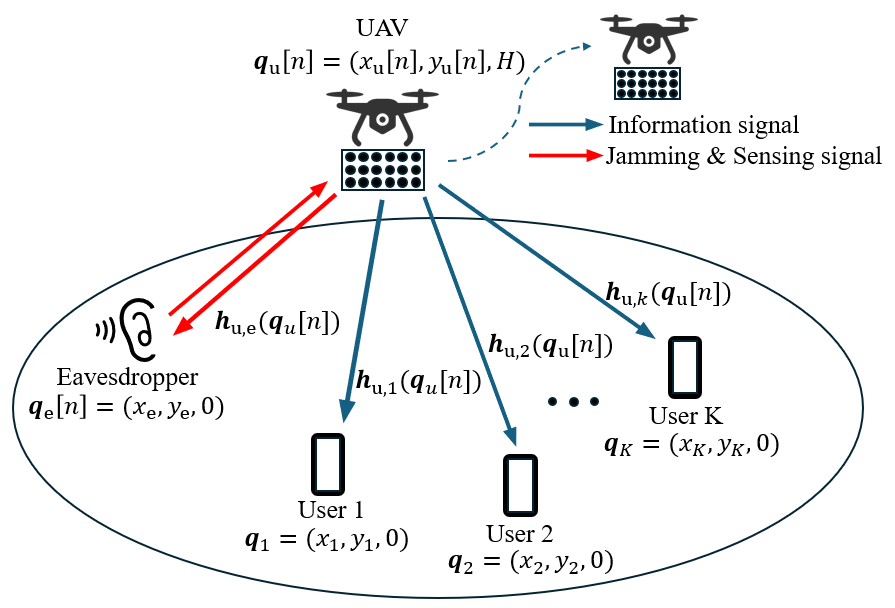}
    \caption{UAV-enabled ISAC system}
    \label{fig:my_label}
\end{figure}
 We consider a UAV-enabled secure ISAC system as shown in Fig. 1, where the full-duplex UAV equipped with UPA acting as base station provides the downlink communications to $K$ single-antenna ground users, while attempting to detect the passive eavesdropper with the echo of the jamming signal reflected to it in the given period $T$. For tractability, the entire mission period $T$ is divided into $N_{t}$ time slots, each duration of which is $ \delta_{t}=T/N_{t}$ to be sufficiently short enough that the UAV's position can be regarded as being unchanged\cite{jeong2017mobile}. We denote the set of ground users as $\mathcal{K}=\{1, \dots, K\}$ and consider a two-dimensional (2D) Cartesian coordinate system. Each ground user $k$ and eavesdropper are located at ground $\pmb{q}_{k}=(x_{k}, y_{k})$ and $\pmb{q}_{\mathrm{E}}=(x_{\mathrm{E}},y_{\mathrm{E}})$, respectively. The UAV is assumed to maintain a fixed altitude $H$ due to the air traffic control purposes \cite{one}, and therefore, its $xy$-plane location can be denoted as $\pmb{q}_{\mathrm{U}}[n]=(x_{\mathrm{U}}[n], y_{\mathrm{U}}[n])$, for $ 1\leq n\leq N_{t}$. The maximum speed of the UAV is denoted by $v_{\text{max}}$.
Without loss of generality, the number of transmit and receive antennas dispatched at UAV is equivalent and is denoted as $M=M_{\text{x}} \times M_{\text{y}}$ with $M_{\text{x}}$ and $M_{\text{y}}$ representing the number of antenna elements along $x$-axis and $y$-axis, which is assumed to be sufficiently large compared to the number of ground nodes. Moreover, each antenna element in x and y-axis is separated by $\lambda/2$, where $\lambda$ is the carrier wavelength. We assume the perfect channel state information (CSI), e.g., the positioning information of nodes can be obtained via global positioning system (GPS) or camera\cite{gps}.
\vspace{-0.32cm}
\subsection{Communication Model}
We consider Rician fading for the air-to-ground channels following\cite{Professor}.
Accordingly, the baseband equivalent channels from UAV to users and eavesdropper can be written as 
\vspace{-0.2cm}
\begin{align}
\label{channel model}
   &\pmb{h}_{\mathrm{U},i}(\pmb{q}_{\mathrm{U}}[n])=\sqrt{\beta_{i}(\pmb{q}_{\mathrm{U}}[n])}\pmb{\chi}_{\mathrm{U},i}[n], 
   \end{align}
   where
   \vspace{-0.2cm}
   \begin{align}
   &\pmb{\chi}_{\mathrm{U},i}[n]=\sqrt{\frac{K_{\text{r}}}{(K_{\text{r}}+1)}}\pmb{\gamma}_{\mathrm{U},i}[n]+\sqrt{\frac{1}{(K_{\text{r}}+1)}}\tilde{\pmb{\gamma}}_{\mathrm{U},i}[n],
\end{align}
for $i \in \mathcal{K} \cup \{\text{E}\}$, with $K_{\text{r}}$ being Rician factor, $\pmb{\gamma}_{\mathrm{U},i}[n] = 
\sqrt{1/M}( [1,\ldots,e^{-j\pi(M_{\text{x}}-1)\cos(\theta_{i}[n])\sin(\phi_{i}[n]))/2}]
\otimes [1,\ldots, e^{-j2\pi (M_{\text{y}}-1) \sin(\theta_{i}[n]) \sin(\phi_{i}[n])/2}] )^T\in\pmb{C}^{M\times 1}$ being line-of-sight (LOS) component, $\tilde{\pmb{\gamma}}_{\mathrm{U},i}[n]\in\pmb{C}^{M\times 1}$ being non-LOS (NLOS) component modeled as a circularly symmetric complex Gaussian (CSCG) random variable, and $\phi_{i}[n]$ and $\theta_{i}[n]$ being the elevation and azimuth angles of departure (AoD) from UAV to users and eavesdropper. Moreover, in \eqref{channel model}$, \beta_{i}(\pmb{q}_{\mathrm{U}}[n])$ is the channel power gains defined as  
$\beta_{i}(\pmb{q}_{\mathrm{U}}[n])=M\beta_0/(H^2 + d_{\mathrm{U},i}(\pmb{q}_{\mathrm{U}}[n]))$,
where  $d_{\mathrm{U},i}(\pmb{q}_{\mathrm{U}}[n])=\Vert \pmb{q}_{\mathrm{U}}[n] - \pmb{q}_{i} \Vert^2$ defines the distance between UAV and user or eavesdropper, and $\beta_0$ represents the channel power gain at a distance 1 m. The Doppler effects induced by UAV's flying are assumed to be compensated at the receiver side\cite{one}. \\
\indent The UAV transmits the information bearing signal $b_{k}[n]$ to user $k$ and the jamming signal $b_{\mathrm{E}}[n]$ to eavesdropper at the $n$th slot, where $b_{k}[n]$ and $b_{\mathrm{E}}[n]$ follow the complex Gaussian normal distribution with zero mean and they are uncorrelated, i.e., $E[b_{k}[n]b_{k'}[n]] = 0$, for $k, k'$ $\in \mathcal{K} \cup \{\text{E}\}$ and $k\neq k'$\cite{one}.
Accordingly, the received signal at user $k$ and eavesdropper can be expressed as
\vspace{-0.3cm}
\begin{subequations}
\begin{align}
\label{eq:user_receive}
&y_{k}[n] = \pmb{h}_{\mathrm{U},k}^H(\pmb{q}_{\mathrm{U}}[n])\pmb{s}[n]+n_{k} =\underbrace{\pmb{h}_{\mathrm{U},k}^H(\pmb{q}_{\mathrm{U}}[n])\pmb{f}_{k}[n]b_{k}[n]}_{\text{desired signal}}+ \notag  \\ \hspace{-0.1cm}&\underbrace{\pmb{h}_{\mathrm{U},k}^H(\pmb{q}_{\mathrm{U}}[n])\sum_{i\neq k} \pmb{f}_{i}[n] b_i[n]}_{\text{inter-user interference}} 
+  \underbrace{\pmb{h}_{\mathrm{U},k}^H(\pmb{q}_{\mathrm{U}}[n])\pmb{f}_{\mathrm{E}}[n]b_{\mathrm{E}}[n]}_{\text{jamming interference}} + n_{k},
\end{align}
\vspace{-0.3cm}
\begin{align}
&y_{\mathrm{E}}[n] = \pmb{h}_{\mathrm{U,E}}^H(\pmb{q}_{\mathrm{U}}[n])\pmb{s}[n]+n_{\mathrm{E}} =\underbrace{\pmb{h}_{\mathrm{U,E}}^H(\pmb{q}_{\mathrm{U}}[n])\pmb{f}_{k}[n]b_{k}[n]}_{\text{desired signal}}+\notag \\   
\hspace{-0.2cm}&\underbrace{\pmb{h}_{\mathrm{U,E}}^H(\pmb{q}_{\mathrm{U}}[n])\sum_{i\neq k} \pmb{f}_{i}[n] b_i[n]}_{\text{inter-user interference}} 
+  \underbrace{\pmb{h}_{\mathrm{U,E}}^H(\pmb{q}_{\mathrm{U}}[n])\pmb{f}_{\mathrm{E}}[n]b_{\mathrm{E}}[n]}_{\text{jamming interference}} + n_{\mathrm{E}},
\end{align}
\end{subequations}
 respectively, where $\pmb{s}[n]=\sum_{k=1}^K\pmb{f}_{k}[n]b_{k}[n]+\pmb{f}_{\mathrm{E}}[n]b_{\mathrm{E}}[n]$ is transmit signal of UAV at time slot $n$, with $\pmb{f}_{k}[n]$ and $\pmb{f}_{\mathrm{E}}[n]$ being transmit beamforming vectors for user $k$'s data and for jamming at UAV, and $n_{k}$ and $n_{\mathrm{E}}$ are the addictive white Gaussian noise (AWGN) with zero mean and variance of $\sigma_{\mathrm{n}}^2$. By following \cite{secrecy}, the achievable secrecy rate of  user $k$ at the $n$th time slot is calculated as 
 \vspace{-0.3cm}
\begin{multline}
\label{secrecy rate}
\hspace{-0.43cm} R_{\text{sec},k}(\pmb{q}_{\mathrm{U}}[n], \pmb{f}[n]) = [ \log_{2}(1 + \Gamma_{k}(\pmb{q}_{\mathrm{U}}[n], \pmb{f}[n]) \\ - 
\log_{2}(1 + \Gamma_{\mathrm{E},k}(\pmb{q}_{\mathrm{U}}[n], \pmb{f}[n]) ]^+,
\end{multline}
where $\pmb{f}[n] =$ $\{\pmb{f}_{i}[n]\}_{i
\in \{\mathcal{K}\}\cup\{\mathrm{E}\}}$, $\Gamma_{k}(\pmb{q}_{\mathrm{U}}[n], \pmb{f}[n])$ and $\Gamma_{\mathrm{E},k}(\pmb{q}_{\mathrm{U}}[n], \pmb{f}[n])$ represent the signal-to-inter-ference-plus-noise ratios (SINRs) of user $k$ and eavesdropper and are given as $\Gamma_{k}(\pmb{q}_{\mathrm{U}}[n], \pmb{f}[n]) = | \pmb{h}_{\mathrm{U},k}^H(\pmb{q}_{\mathrm{U}}[n])\pmb{f}_{k}[n]|^2 / (\sum_{i\neq k} | \pmb{h}_{\mathrm{U},k}^H(\pmb{q}_{\mathrm{U}}[n]) \pmb{f}_{i}[n] |^2$ $+ | \pmb{h}_{\mathrm{U},k}^H(\pmb{q}_{\mathrm{U}}[n]) \pmb{f}_{\mathrm{E}}[n]|^2 + \sigma_{\text{n}}^2)$ and $\Gamma_{\text{E},k}(\pmb{q}_{\mathrm{U}}[n], \pmb{f}[n]) =  |\pmb{h}_{\mathrm{U,E}}^H(\pmb{q}_{\mathrm{U}}[n]) \pmb{f}_{k}[n]|^2/(\sum_{i\neq k}  |\pmb{h}_{\mathrm{U,E}}^H(\pmb{q}_{\mathrm{U}}[n]) \pmb{f}_{i}[n]|^2 $ $+ | \pmb{h}_{\mathrm{U,E}}^H(\pmb{q}_{\mathrm{U}}[n]) \pmb{f}_{\mathrm{E}}[n] |^2 + \sigma_{\text{n}}^2)$,
respectively.
\vspace{-0.3cm}
\subsection{Sensing Model}
For defining the sensing model, we use the reciprocity between uplink and downlink channels\cite{sensing}, and the reflected signals to pass through two or more paths are assumed to be neglected.  The received sensing echo at the UAV is then given as 
\vspace{-0.3cm}
\begin{align}
\label{sensing model}
    &\pmb{y}_{\mathrm{s}}[n] = \underbrace{\zeta_{\mathrm{U,E}}(\pmb{q}_{\mathrm{U}}[n])\pmb{\chi}_{\mathrm{U,E}}[n] \pmb{\chi}_{\mathrm{U,E}}^H[n] \pmb{s}[n]}_{\text{sensing echo from eavesdropper}}  \\
    & \underbrace{\sum_{{k}=1}^K \zeta_{\mathrm{U},k}(\pmb{q}_{\mathrm{U}}[n]) \pmb{\chi}_{\mathrm{U},k}[n] \pmb{\chi}_{\mathrm{U},k}^H[n]  \pmb{s}[n]}_{\text{sensing echos from users}} + \underbrace{\pmb{H}_{\mathrm{S,I}}[n] \pmb{s}[n]}_{\text{self-interference}} + \pmb{n}_{\mathrm{s}}[n], \nonumber
\end{align}
where $\pmb{n}_{\mathrm{s}}[n]$ represents an AWGN vector following $\mathcal{CN}(0, \sigma_{\text{n}}^2 \pmb{I}_{M})$,  $\pmb{H}_{\mathrm{S,I}}[n]\in\pmb{C}^{M \times M}$ is the self-interference matrix due to the FD operation at UAV,
and $\zeta_{\mathrm{U},i}(\pmb{q}_{\mathrm{U}}[n])$, for $i \in \mathcal{K} \cup \{\text{E}\}$, is the power gain between UAV and user $k$ and eavesdropper and is defined as 
$\zeta_{\mathrm{U},i}(\pmb{q}_{\mathrm{U}}[n])=\sqrt{M^2\beta_0\rho_0/d_{\mathrm{U},i}(\pmb{q}_{\mathrm{U}}[n])^2}$,\cite{sensing}
with $\rho_0$ being the eavesdropper's radio cross section (RCS). Since the UAV knows its own transmit signal, the sensing echos reflected by the users in \eqref{sensing model} are assumed to be perfectly canceled\cite{sensing}. For improving the sensing accuracy at UAV, the receive filtering vector $\pmb{w}[n]$ is adopted. The SINR of the sensing echo at UAV in time slot $n$ can be then represented as
\vspace{-0.2cm}
\begin{align}
\label{sensing SINR}
    &\Gamma_{\mathrm{s}}(\pmb{q}_{\mathrm{U}}[n], \pmb{f}[n], \pmb{w}[n]) =  \\ 
    &\frac{\sum_{{i}=1}^K\left|\pmb{a}^H(\pmb{q}_{\mathrm{U}}[n], \pmb{w}[n])\pmb{f}_{i}[n]\right|^2+\left|\pmb{a}^H(\pmb{q}_{\mathrm{U}}[n], \pmb{w}[n])\pmb{f}_{\mathrm{E}}[n]\right|^2}{\sum_{{i}=1}^K \left| \pmb{w}^H[n] \pmb{H}_{\mathrm{S,I}}[n]\pmb{f}_{i}[n]\right|^2+ \left| \pmb{w}^H[n] \pmb{H}_{\mathrm{S,I}}[n]\pmb{f}_{\mathrm{E}}[n]\right|^2+\sigma_{\text{n}}^2} \notag,
\end{align} 
where $\pmb{a}^H(\pmb{q}_{\mathrm{U}}[n], \pmb{w}[n])=\zeta_{\mathrm{U,E}}(\pmb{q}_{\mathrm{U}}[n]) \pmb{w}^H[n] \pmb{\chi}_{\mathrm{U,E}}[n] \pmb{\chi}_{\mathrm{U,E}}^H[n]$. The detection of eavesdropper's presence can be effectively achieved, e.g., by using Kalman filter or matched filter, given that the sensing SINR $\Gamma_{\mathrm{s}}(\pmb{q}_{\mathrm{U}}[n], \pmb{f}[n], \pmb{w}[n])$ in \eqref{sensing SINR} is strong enough to be larger than the predetermined threshold.
\vspace{-0.3cm}
\section{Problem formulation and proposed algorithm}
In this work, we aim at maximizing the sum secrecy rates of downlink users by jointly optimizing transceiver beamforming and UAV’s trajectory subject to the constraints of sensing SINR, and UAV's operational capability. To this end, the problem is formulated as
\vspace{-0.2cm}
\begin{subequations}
\label{original}
\begin{align}
& \underset{\pmb{f}[n], \pmb{q}_{\mathrm{U}}[n], \pmb{w}[n]}{\text{max}}
\sum_{n=1}^{N_{t}} \sum_{k=1}^K R_{\mathrm{sec},k}(\pmb{q}_{\mathrm{U}}[n], \pmb{f}[n]) \label{eq:original objective},\\
& \text{s.t.} \sum_{{i}=1}^K\| \pmb{f}_{i}[n] \|^2 + \| \pmb{f}_{\mathrm{E}}[n] \|^2 \leq P_{\text{max}}, \forall n, \label{eq:original power const.}\\
&\quad \| \pmb{q}_{\mathrm{U}}[n] - \pmb{q}_{\mathrm{U}}[n-1] \|^2 \leq \delta_{t} v_{\text{max}},  \forall n, \label{eq:original velocity const.}\\
&\quad \pmb{q}_{\mathrm{U}}[1] = \pmb{q}_{\mathrm{0}}, \quad \pmb{q}_{\mathrm{U}}[N] = \pmb{q}_{\mathrm{f}}, \quad  \label{eq:original position const.}\\
&\quad \| \pmb{w}[n] \|^2 = 1,   \forall n \label{eq:original receive beamforming const.}, \\
&\quad \Gamma_{\mathrm{s}}(\pmb{q}_{\mathrm{U}}[n],\pmb{f}[n],\pmb{w}[n]) \geq \Gamma_{\text{th}},  \forall n, \label{eq:original sensing SINR const.}
\end{align}
\end{subequations}
where $\pmb{q}_{0}$ and $\pmb{q}_{\mathrm{f}}$ are the initial and final positions of the UAV. The constraint \eqref{eq:original power const.} and  \eqref{eq:original velocity const.} represent the transmit power and maximum velocity constraint for the UAV, respectively, \eqref{eq:original position const.} and \eqref{eq:original receive beamforming const.} are constraints for the UAV's initial and final position and for the normalized receive beamforming, and \eqref{eq:original sensing SINR const.} represents the sensing SINR constraint for satisfying the required sensing accuracy.
The problem \eqref{original} is non-convex due to the non-convexity of \eqref{eq:original objective}, \eqref{eq:original power const.}, \eqref{eq:original receive beamforming const.} and \eqref{eq:original sensing SINR const.}, whose solutions can be obtained via Algorithm 1 based on BCD method \cite{BCD} and SDP relaxation technique\cite{convexopt}, and the details are provided in the following. 
\vspace{-0.45cm}
\subsection{Optimization of Receive Beamforming}
In \eqref{original}, the receive beamforming vector $\pmb{w}[n]$ only affects to the sensing SINR constraint \eqref{eq:original sensing SINR const.} under the contraint \eqref{eq:original receive beamforming const.}. To this end, we design $\pmb{w}[n]$ to maximize the sensing SINR $\Gamma_{\mathrm{s}}(\pmb{q}_{\mathrm{U}}[n],\pmb{f}[n],\pmb{w}[n])$, as 
\vspace{-0.2cm}
\begin{align}
\label{opt_w[n]}
&\pmb{w}_{\text{opt}}[n] = 
\frac{A(\pmb{f}[n])^{-1} \pmb{\chi}_{\mathrm{U,E}}[n] \pmb{\chi}_{\mathrm{U,E}}^H[n](\sum_{{i}=1}^K\pmb{f}_{i}[n] + \pmb{f}_{\mathrm{E}}[n])}{\| A(\pmb{f}[n])^{-1} \pmb{\chi}_{\mathrm{U,E}}[n] \pmb{\chi}_{\mathrm{U,E}}^H[n](\sum_{{i}=1}^K\pmb{f}_{i}[n] +  \pmb{f}_{\mathrm{E}}[n]) \|},
\end{align} 
where $A(\pmb{f}[n])$$=$$\pmb{H}_{\mathrm{S,I}}[n]$$(\sum_{{i}=1}^K\pmb{f}_{i}[n]\pmb{f}_{i}^H[n]$$ + $$ \pmb{f}_{\mathrm{E}}[n]\pmb{f}_{\mathrm{E}}^H[n] )\\ \pmb{H}_{\mathrm{S,I}}^{H}[n]$$ +\sigma_{\text{n}}^2\pmb{I}_{M}$.
\vspace{-0.45cm}
\subsection{Optimization of UAV's Trajectory}
Here, we develop the UAV's trajectory $\pmb{q}_{\mathrm{U}}[n]$, for all $n$. To this end, we introduce slack variables $\alpha_{k}[n]$ and $\gamma[n]$ for satisfying 
\vspace{-0.2cm}
\begin{subequations}
\begin{align}
&\alpha_{k}[n] \geq H^2 + \| \pmb{q}_{\mathrm{U}}[n] - \pmb{q}_{k} \|^2, \forall n, \forall k,\label{eq:slack_x const.q} \\
&\gamma[n] \leq H^2 + \| \pmb{q}_{\mathrm{U}}[n] - \pmb{q}_{\mathrm{E}} \|^2,  \forall n,\label{eq:slack_y const.q} 
\end{align}
\end{subequations}
and reformulate \eqref{original} as
\begin{subequations}
\label{slack_variable_q}
\begin{align}
\max_{\pmb{q}_{\mathrm{U}[n]},\alpha_{k}[n],\gamma[n]}& 
 \sum_{n=1}^{N_{t}} \sum_{k=1}^K \left[ \log_{2}\left(1 + \frac{c_{k}[n]}{C_{k}[n] - c_{k}[n] + C_{\beta} \alpha_{k}[n]}\right) \right. \notag \\ 
& \left. - \log_{2}\left(1 + \frac{c_{\mathrm{E},k}[n]}{C_{E}[n] - c_{\mathrm{E},k}[n] + C_{\beta} \gamma[n]}\right) \right] \label{eq:objective_q_refine}, \\
\text{s.t.} \quad
& \eqref{eq:original velocity const.}, \eqref{eq:original position const.}, \eqref{eq:original sensing SINR const.}, \eqref{eq:slack_x const.q}, \eqref{eq:slack_y const.q} \label{q_constraints},
\end{align}
\end{subequations}
where ${c_{k}[n]}=| \pmb{\chi}_{\mathrm{U},k}^H[n]\pmb{f}_{k}[n] |^2$, ${c_{\mathrm{E},k}[n]}=\| \pmb{\chi}_{\mathrm{U,E}}^H[n]\pmb{f}_{k}[n] \|^2$, ${C_{k}[n]}=\sum_{{i}=1}^K \| \pmb{\chi}_{\mathrm{U},k}^H[n]\pmb{f}_{i}[n] \|^2 + \| \pmb{\chi}_{\mathrm{U},k}^H[n]\pmb{f}_{\mathrm{E}}[n] \|^2$,  ${C_{\mathrm{E}}[n]}=\sum_{{i}=1}^K \| \pmb{\chi}_{\mathrm{U,E}}^H[n]\pmb{f}_{i}[n] \|^2 + \| \pmb{\chi}_{\mathrm{U,E}}^H[n]\pmb{f}_{\mathrm{E}}[n] \|^2$, $C_{\beta}=\sigma_{\text{n}}^2/M\beta_0$.
To resolve the non-convexity of \eqref{slack_variable_q}, we apply the first-order Taylor expansion to \eqref{eq:objective_q_refine} and \eqref{eq:slack_y const.q}, yielding
\vspace{-0.2cm}
\begin{subequations}
\label{slack_variable_q_taylor}
\begin{align}
&\max_{\substack{\pmb{q}_{\mathrm{U}}[n],\alpha_{k}[n],\gamma[n]}} \sum_{n=1}^{N_{t}} \sum_{k=1}^{K} \log_{2}\left(C_{k}[n]+C_{\beta} \alpha_{k}[n]\right) - \notag \\
&\quad \quad \quad \quad \frac{C_{\beta}(\alpha_{k}[n]-\tilde{\alpha_{k}}[n])}{\ln{2}\; (C_{k}[n]-c_{k}[n]+C_{\beta} \tilde{\alpha_{k}}[n])}-\notag \\
&\quad \quad \quad \quad \log_{2}\left(C_{k}[n]-c_{k}[n]+C_{\beta} \tilde{\alpha_{k}}[n]\right)- \notag \\
&\quad \quad \quad \quad \log_{2}(1+\frac{c_{\mathrm{E},k}[n]}{C_{E}[n] - c_{\mathrm{E},k}[n] +  C_{\beta}\gamma[n]}),  \label{eq:objective_q_taylor}  \\
&\text{s.t.} \quad \gamma[n] \leq H^2 +2(\tilde{\pmb{q}}_{\mathrm{U}}[n] - \pmb{q}_{\mathrm{E}} )^T (\pmb{q}_{\mathrm{U}}[n] - \tilde{\pmb{q}}_{\mathrm{U}}[n]) \nonumber \\
& + \| \tilde{\pmb{q}}_{\mathrm{U}}[n] - \pmb{q}_{\mathrm{E}} \|^2,  \forall n, \label{eq:slack_y const.qtaylor} \\
& \eqref{eq:slack_x const.q}, \eqref{q_constraints}, \nonumber
\end{align}
\end{subequations}
where $\tilde{\alpha_{k}}[n]$ and $\tilde{\gamma}[n]$ are feasible values of $\alpha_{k}[n]$ and $\gamma[n]$. Since the problem \eqref{slack_variable_q_taylor} is convex, it can be solved by using CVX\cite{cvx}.
\vspace{-0.3cm}
\subsection{Optimization of Transmit Beamforming}
Finally, we optimize the transmit beamforming vectors $\pmb{f}[n]$, for which we have
\vspace{-0.1cm}
\begin{align}
\label{optimization_f_i}
\underset{\pmb{f}[n]}{\text{max}} \sum_{n=1}^{N_{t}} \sum_{k=1}^K R_{\text{sec},k}(\pmb{f}[n]), \quad \text{s.t. }  \eqref{eq:original power const.},\eqref{eq:original sensing SINR const.}.
\end{align}
Since $R_{\text{sec},k}(\pmb{f}[n])$ and constraints of \eqref{optimization_f_i} are non-convex owing to quadratic form, we adopt the SDP relaxation \cite{sixteen}.
Specifically, by defining the beamforming covariance matrices $\pmb{F}_{k}[n] = \pmb{f}_{k}[n]\pmb{f}_{k}[n]^H$ and $\pmb{F}_{\mathrm{E}}[n] = \pmb{f}_{\mathrm{E}}[n]\pmb{f}_{\mathrm{E}}[n]^H$, we can rewrite \eqref{optimization_f_i} in terms of $\pmb{F}_{k}[n]$ and $\pmb{F}_{\mathrm{E}}[n]$ with adding the rank-1 constraint and their positive semi-definiteness constraint as
\vspace{-0.4cm}
\begin{subequations}\label{optimization_F_i}
\begin{align}
    \max_{\pmb{F}[n]} & \sum_{n=1}^{N_{t}} \sum_{k=1}^K R_{\text{sec},k}(\pmb{F}[n]), & \label{objective F} \\
    \text{s.t.} \quad & \sum_{i=1}^K \text{Tr}(\pmb{F}_{i}[n]) + \text{Tr}(\pmb{F}_{\mathrm{E}}[n]) \leq P_{\text{max}}, &&\forall n, \label{F power const} \\
    & \Gamma_{\mathrm{s}}(\pmb{F}[n]) \geq \Gamma_{\text{th}}, &&\forall n, \label{sensing threshold F_i} \\
    & \pmb{F}_{i}[n] \succeq 0, \pmb{F}_{\mathrm{E}}[n] \succeq 0, \quad \quad \quad \quad \forall i, &&\forall n, \label{semidefinite const} \\
    & \text{rank}(\pmb{F}_{i}[n]) = \text{rank}(\pmb{F}_{\mathrm{E}}[n]) = 1, \forall i, &&\forall n, \label{rank constraint}
\end{align}
\end{subequations}
where $\pmb{F}[n] =$ $\{\pmb{F}_{i}[n]\}_{i\in \{\mathcal{K}\}\cup\{\mathrm{E}\}}$. Similar with Sec. III-B, the first-order Taylor \cite{Professor} is applied to \eqref{objective F}, which provides 
\vspace{-0.4cm}
\begin{align}
\label{R_approx_F}
&\hat{R}_{\text{sec},k}(\pmb{F}[n]) =  \\
&\log_{2}\left(\pmb{h}_{\mathrm{U},k}^H[n](\sum_{{i}=1}^K\pmb{F}_{i}[n]+\pmb{F}_{\mathrm{E}}[n]) \pmb{h}_{\mathrm{U},k}[n] + \sigma_{\text{n}}^2\right) - D_{\mathrm{U},k}[n] \notag
\end{align}
\vspace{-0.4cm}
\begin{align}
&-\text{Tr}\left(\frac{\pmb{H}_{\mathrm{U},k}[n](\sum_{i\neq k}\pmb{F}_{i}[n]-\pmb{\tilde{F}}_{i}[n]+\pmb{F}_{\mathrm{E}}[n]-\pmb{\tilde{F}}_{\mathrm{E}}[n])}{\ln{2}\;(\pmb{h}_{\mathrm{U},k}^H[n](\sum_{i\neq k}\pmb{\tilde{F}}_{i}[n]+\pmb{\tilde{F}}_{\mathrm{E}}[n]) \pmb{h}_{\mathrm{U},k}[n] + \sigma_{\text{n}}^2)}\right)  \nonumber
\end{align}
\vspace{-0.4cm}
\begin{align}
& -\text{Tr}\left(\frac{\pmb{H}_{\mathrm{U,E}}[n](\sum_{{i}=1}^{K}\pmb{F}_{i}[n]-\pmb{\tilde{F}}_{i}[n]+\pmb{F}_{\mathrm{E}}[n]-\pmb{\tilde{F}}_{\mathrm{E}}[n])}{\ln{2}\;(\pmb{h}_{\mathrm{U,E}}^H[n](\sum_{{i}=1}^{K}\pmb{\tilde{F}}_{i}[n]+\pmb{\tilde{F}}_{\mathrm{E}}[n]) \pmb{h}_{\mathrm{U,E}}[n] + \sigma_{\text{n}}^2)}\right) \nonumber 
\end{align}
\vspace{-0.4cm}
\begin{align}
-D_{\mathrm{U,E}}[n]+\log_{2}\left(\pmb{h}_{\mathrm{U,E}}^H(\sum_{\text{i} \neq \text{k}}\pmb{F}_{i}[n]+\pmb{F}_{\mathrm{E}}[n]) \pmb{h}_{\mathrm{U,E}}[n] + \sigma_{\text{n}}^2\right), \nonumber 
\end{align}
where $\pmb{\tilde{F}}_{i}[n]$ and $\pmb{\tilde{F}}_{\mathrm{E}}[n]$ are feasible values of $\pmb{F}_{i}[n]$ and $\pmb{F}_{\mathrm{E}}[n]$,\\
$D_{\mathrm{U},k}[n]=\log_{2}(\pmb{h}_{\mathrm{U},k}^H[n](\sum_{i\neq k}\tilde{\pmb{F}}_{i}[n]+\tilde{\pmb{F}}_{\mathrm{E}}[n]) \pmb{h}_{\mathrm{U},k}[n] + \sigma_{\text{n}}^2)$ \\ \hspace{-0.1cm} $D_{\mathrm{U,E}}[n]=\log_{2}(\pmb{h}_{\mathrm{U,E}}^H[n](\sum_{{i}=1}^{K}\tilde{\pmb{F}}_{i}[n]+\tilde{\pmb{F}}_{\mathrm{E}}[n]) \pmb{h}_{\mathrm{U,E}}[n]\hspace{-0.05cm}+ \hspace{-0.05cm} \sigma_{\text{n}}^2)$. By the denominator of $\Gamma_{\mathrm{s}}(\pmb{F}[n])$ is nonzero, equation \eqref{sensing threshold F_i} can be rewritten as
\begin{align}
\label{F constraint convex form}
&\Gamma_{\text{th}} \biggl(\pmb{w}^H[n]\pmb{H}_{\mathrm{S,I}}[n] \sum_{{i}=1}^K\left(\pmb{F}_{i}[n]+\pmb{F}_{\mathrm{E}}[n]\right) \pmb{H}_{\mathrm{S,I}}^H[n]\pmb{w}[n]  + \sigma_{\text{n}}^2 \biggl) \notag \\
&-\biggl( \pmb{a}^H[n] \left( \sum_{{i}=1}^K \pmb{F}_{i}[n] + \pmb{F}_{\mathrm{E}}[n] \right) \pmb{a}[n]\biggl)\leq 0,
\end{align} 
where $\pmb{a}^H[n]=\zeta_{\mathrm{U,E}}(\pmb{q}_{\mathrm{U}}[n]) \pmb{w}^H[n] \pmb{\chi}_{\mathrm{U,E}}[n] \pmb{\chi}_{\mathrm{U,E}}^H[n]$. With \eqref{R_approx_F} and \eqref{F constraint convex form} and by relaxing \eqref{rank constraint}, we have the convex problem to design the covariance matrices $\pmb{F}_{k}[n]$ and $\pmb{F}_{\mathrm{E}}[n]$, which can be devloped via CVX\cite{cvx}. By following the rank-1 extraction \cite{convexopt}, we design the optimal transmit beamforming vector as  $\pmb{f}_{k}[n]=\rho_{k}[n]\sqrt{\lambda_{\mathrm{max}}(\pmb{F}_{k}[n])}\pmb{v}_{\mathrm{max}}(\pmb{F}_{k}[n])$ and $\pmb{f}_{\mathrm{E}}[n]=\rho_{\mathrm{E}}[n]\sqrt{\lambda_{\mathrm{max}}(\pmb{F}_{\mathrm{E}}[n])}\pmb{v}_{\mathrm{max}}(\pmb{F}_{\mathrm{E}}[n])$, where $\rho_{k}[n]$ and $\rho_{\mathrm{E}}[n]$ are the scaling factor for feasibility. 
\vspace{-0.2cm}
\subsection{Convergence and Complexity Analysis}
\begin{algorithm}[t]
\caption{Algorithm for the \eqref{original}}\label{algorithm}

\begin{algorithmic}[1]

\STATE \textbf{Initialization:} Set $l=0$,  $R_{\text{sec}}^{(0)}=0$, $\epsilon$, and initialize
$\pmb{q}_{\mathrm{U}}^{(l)}$,  $\pmb{\alpha}_{k}^{(l)}$, $\pmb{\gamma}^{(l)}$, $\pmb{w}^{(l)}$, $\pmb{f}_{k}^{(l)}$ and $\pmb{f}_{\mathrm{E}}^{(l)}$.
\WHILE{$\lvert(R_{\text{sec}}^{(l)} -R_{\text{sec}}^{(l-1)}/R_{\text{sec}}^{(l)} \rvert> \epsilon$}
\STATE $l = l + 1$
\STATE Update $\pmb{q}_{\mathrm{U}}^{(l)}, \alpha_{k}^{(l)}$, $\gamma^{(l)}$, via \eqref{slack_variable_q_taylor} with ($\pmb{q}_{\mathrm{U}}^{(l-1)}$, $\alpha_{k}^{(l-1)}$, $\gamma^{(l-1)}$, $\pmb{f}_{k}^{(l-1)}$, $\pmb{f}_{\mathrm{E}}^{(l-1)}$ and $\pmb{w}^{(l-1)}$).
\STATE Update $\pmb{w}^{(l)}$ via \eqref{opt_w[n]} with ($\pmb{f}_{k}^{(l-1)}$, $\pmb{f}_{\mathrm{E}}^{(l-1)}$ and $\pmb{q}_{\mathrm{U}}^{(l)}$).
\STATE Solve \eqref{optimization_F_i} with ($\pmb{q}_{\mathrm{U}}^{(l)}$, $\pmb{w}^{(l)}$, $\pmb{f}_{k}^{(l-1)}$, $\pmb{f}_{\mathrm{E}}^{(l-1)}$)
\STATE Update $\pmb{f}_{k}^{(l)}[n] \leftarrow \rho_{k}[n]\sqrt{\lambda_{\mathrm{max}}(\pmb{F}_{k}[n])}\pmb{v}_{\mathrm{max}}(\pmb{F}_{k}[n])$, $\pmb{f}_{\mathrm{E}}^{(l)}[n] \leftarrow \rho_{\mathrm{E}}[n]\sqrt{\lambda_{\mathrm{max}}(\pmb{F}_{\mathrm{E}}[n])}\pmb{v}_{\mathrm{max}}(\pmb{F}_{\mathrm{E}}[n])$.
\STATE Calculate $R_{\text{sec}}^{(l)}$($\pmb{f}_{k}^{(l)}$, $\pmb{f}_{\mathrm{E}}^{(l)}$, $\pmb{q}_{\mathrm{U}}^{(l)}$)
\ENDWHILE
\end{algorithmic}
\end{algorithm}
\textit{1) Convergence Analysis:} Since the overall Algorithm 1 is based on BCD method, in order to guarantee its convergence, the UAV's trajectory $\pmb{q}_{\mathrm{U}}[n]$ and transmit beamforming $\pmb{f}[n]$ at each iteration need to be optimally and accurately designed. The problem \eqref{slack_variable_q_taylor} and \eqref{optimization_f_i} for the step 4 and 7 of Algorithm 1 are based on the first-order Taylor expansion to guarantee convergence, as they are optimally solved, while ensuring increasing achievable secrecy rate \cite{convergence}. Moreover, the receive beamforming vector $\pmb{w}[n]$ is designed as in the closed-form, by which the proposed Algorithm 1 can converge to the locally optimal solution of the problem \eqref{original}. \\
\indent \textit{2) Complexity Analysis:} In each iteration of Algorithm 1, the interior-point method is used with being coupled with CVX. Particularly, the computational complexity of UAV's trajectory design \eqref{slack_variable_q_taylor} is calculated as $\mathcal{O}((KN_{\text{t}})^{3.5})$ \cite{Professor}, and the complexity of transmit beamforming design \eqref{optimization_F_i} is derived as $\mathcal{O}((KM^2N_{\text{t}})^{3.5})$. The computational complexity of receive beamforming design \eqref{opt_w[n]} is $\mathcal{O}(M^{3})$ due to the matrix inversion. Consequently, the total computational complexity of Algorithm 1 is calculated by $\mathcal{O}(I_{1}(I_{2}(KM^2N_{\text{t}})^{3.5} + M^{3}))$, with $I_{1}$ and $I_{2}$ being the number of iterations required for the step 3 to 8 and for the step 6 to 8, respectively. \vspace{-0.3cm}
\section{Numerical results}
\begin{figure}[t]
    \centering
    \includegraphics[width=7.7cm]{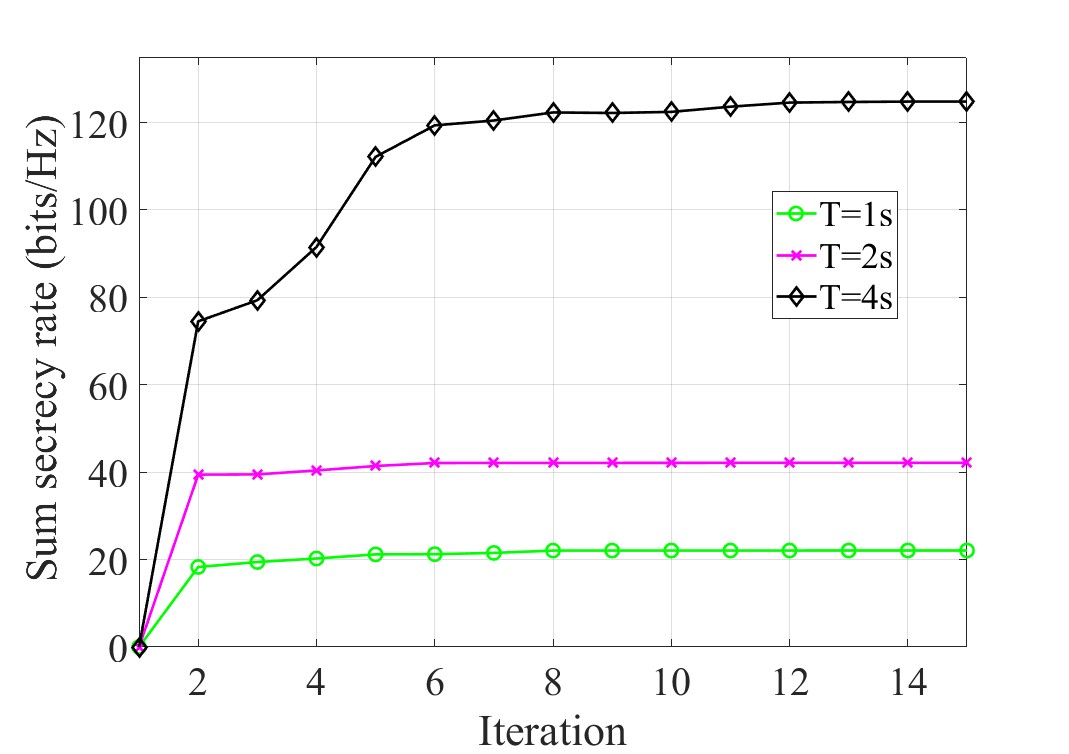}
    \caption{Convergence analysis of Algorithm 1 ($\Gamma_{\text{th}}=10$dB and $P_{\text{max}}= 5W$)}
    \label{fig:enter-label}
\end{figure}
\begin{figure}
    \centering
    \begin{subfigure}{.5\textwidth} 
        \centering
        \includegraphics[width=6.7cm]{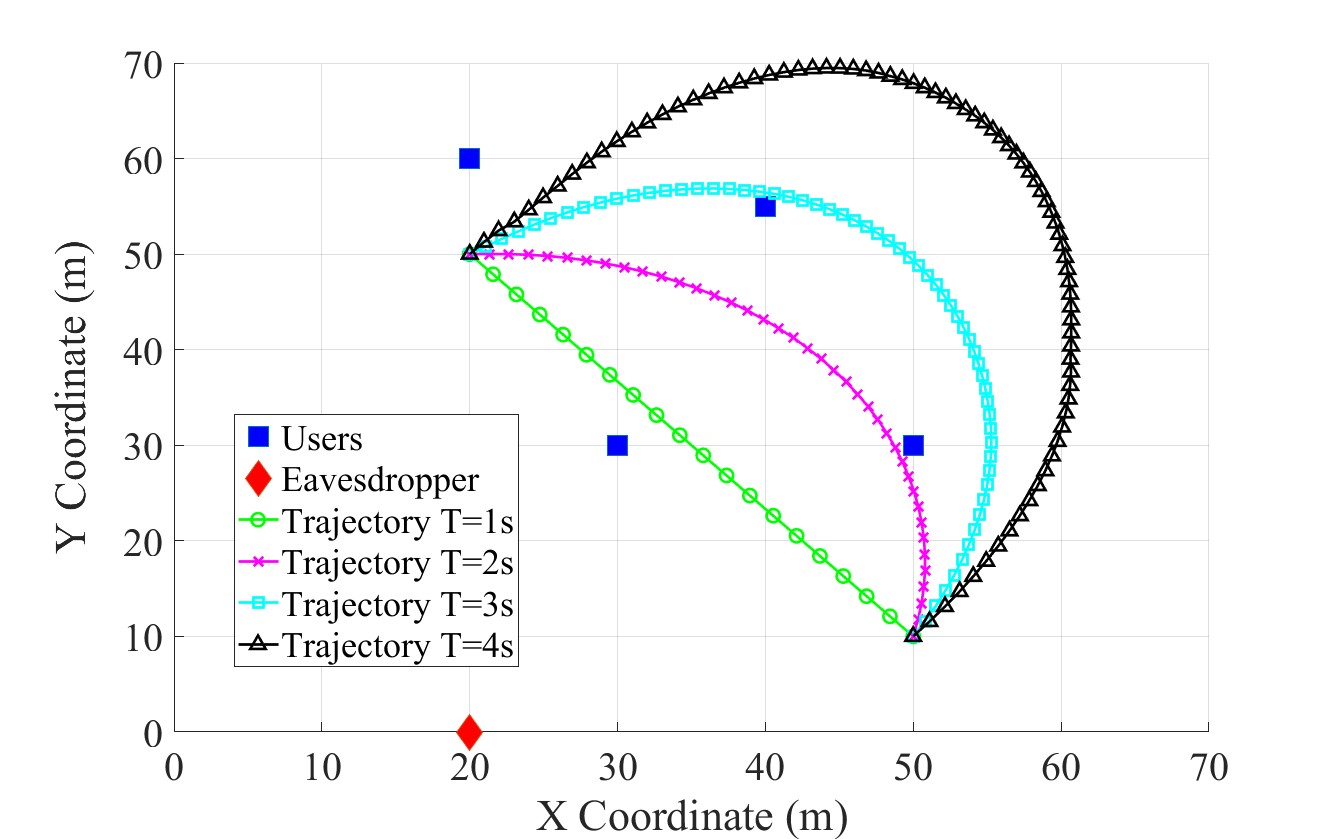}
        \subcaption{$\Gamma_{\text{th}}=\text{10}$dB}
        \label{fig:trajectory}
    \end{subfigure}%
    \\
    \begin{subfigure}{.5\textwidth} 
        \centering
        \includegraphics[width=6.7cm]{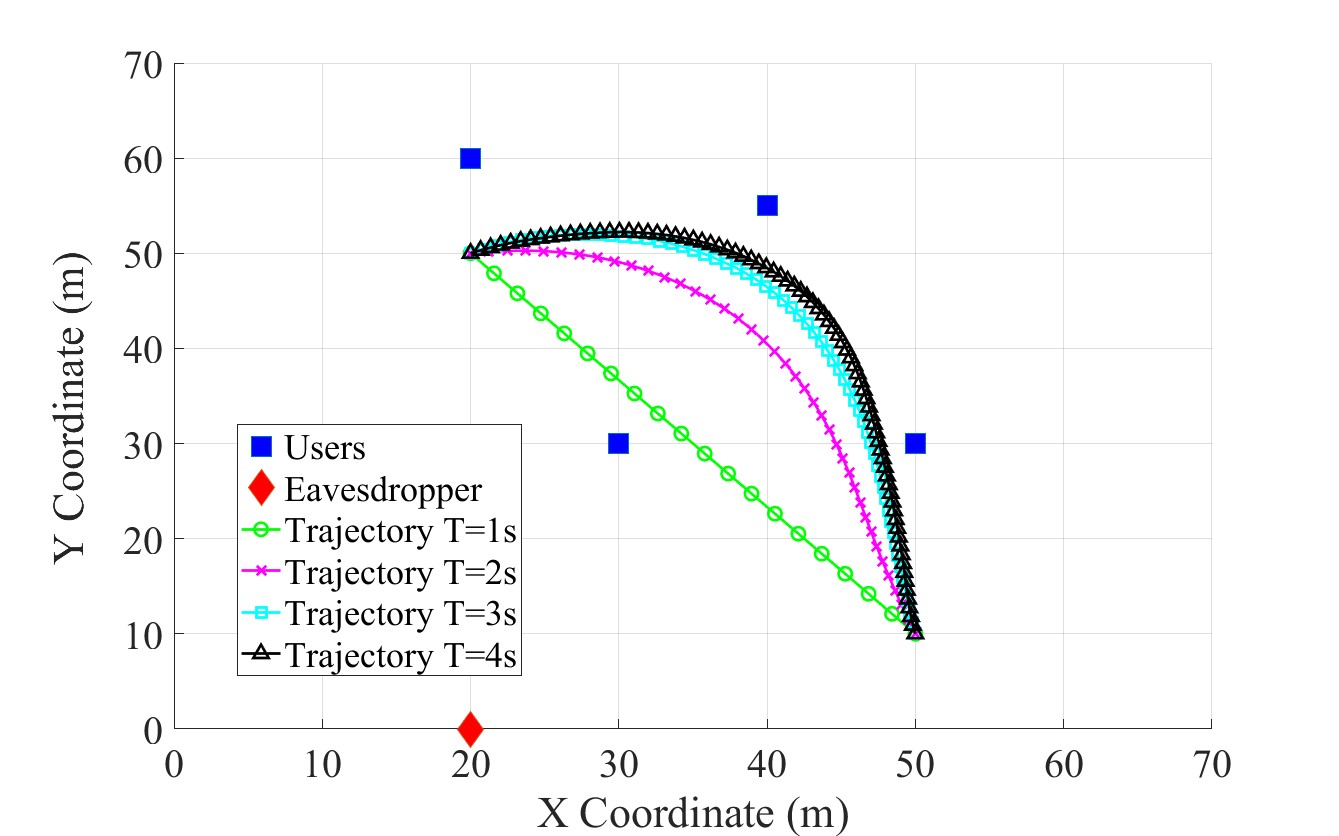}
        \subcaption{$\Gamma_{\text{th}}=\text{30}$dB}
        \label{fig:sensing}
    \end{subfigure}
    \caption{Optimal UAV's trajectory according to different mission periods ($P_{\text{max}}= 5W$): (a) $\Gamma_{\text{th}}=10$dB, (b) $\Gamma_{\text{th}}=30$dB.}
    \label{fig:trajectories}
\end{figure}
In this section, we present the numerical results to evaluate the performance of our proposed Algorithm compared to the reference schemes. For reference, we consider the following schemes: \textit{i) No trajectory. opt.}\cite{ten} to optimize the transmit beamforming vectors and receive beamforming vector by Algorithm 1 with the fixed UAV's trajectory to fly straight from $\pmb{q}_{\mathrm{0}}$ to $\pmb{q}_{\mathrm{F}}$ with the same velocity, and \textit{ii) No transmit beamforming. opt.}\cite{seven} to optimize the UAV's trajectory and receive beamforming by Algorithm 1 with the fixed transmit beamforming of $\pmb{f}_{k}[n]=\sqrt{(P_{\text{max}}/(K+1)}\pmb{h}_{\mathrm{U},k}[n]/\left\|\pmb{h}_{\mathrm{U},k}[n]\right\|$ and $\pmb{f}_{\mathrm{E}}[n]=\sqrt{(P_{\text{max}}/(K+1)}\pmb{h}_{\mathrm{U},\mathrm{E}}[n]/\left\|\pmb{h}_{\mathrm{U},\mathrm{E}}[n]\right\|$. By following the recent UAV-enabled ISAC works\cite{one}, \cite{ten}, we set $H=40$m, $\delta_{t}=0.05s$, $M_{\mathrm{x}}=M_{\mathrm{y}}=3$, $v_{\mathrm{max}}=50$m/s, $\sigma_{\text{n}}^2=-110$dbm , $K_{\mathrm{r}}=500$ and $\beta_{\text{0}}=-30$dB. We also consider the squared area of 70 m $\times$ 70 m, where the $4$ users and eavesdropper are located at $\pmb{q}_{\text{1}}=(20, 60)$, $\pmb{q}_{\text{2}}=(30, 30)$, $\pmb{q}_{\text{3}}=(40, 55)$,  $\pmb{q}_{\text{4}}=(50, 30)$ and $\pmb{q}_{\mathrm{E}}=(20, 0)$, respectively, while the initial and final position of UAV set as $\pmb{q}_{\text{0}}=(20,50)$ and $\pmb{q}_{\text{f}}=(50,10)$.
\begin{figure}[t]
    \centering
    \includegraphics[width=8cm]{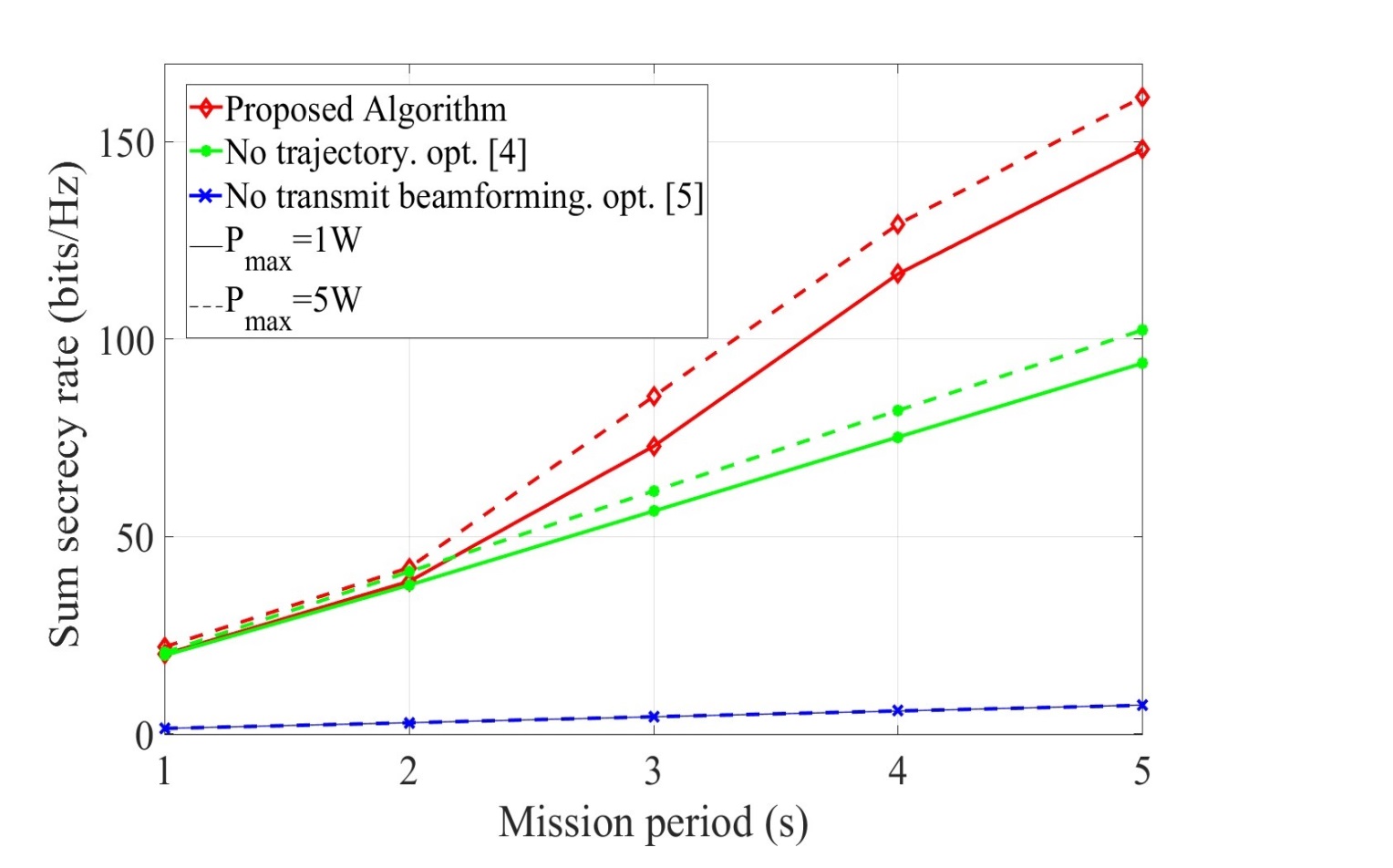}
    \caption{Sum secrecy rate versus the mission time $T$ ($\Gamma_{\text{th}}=10$dB).}
    \label{fig:enter-label}
\end{figure}
\\ \indent Fig. 2 illustrates the convergence of the proposed Algorithm 1. It is verified that the proposed algorithm guarantees the convergence within 10 iterations for the differnt mission period, where the convergence speed becomes faster with the smaller the mission time $T$. Fig. 3 shows the optimal UAV's trajectory obtained by Algorithm 1 with the different mission period and the sensing constraints. In Fig. 3, the mission period becomes shorter, the UAV's trajectory is limited by the initial and final position requirements and, therefore, cannot be sufficiently optimized so as to prevent information leakage to eavesdropper. On the other hand, as the mission period $T$ increases, the UAV's flying paths can be designed to move away from the eavesdropper. Similarly, the higher SINR constraint for sensing in Fig. 3 (b) limits the UAV's mobility compared to Fig. 3 (a) owing to the loss of secrecy rate for satisfying the sensing requirement.\\
 \indent Fig. 4, shows the sum secrecy rates of ground users versus the mission time $T$  compared to the benchmark schemes. It is noticeable that the proposed algorithm provides the best performance in terms of sum secrecy rate compared to reference schemes. In addition, the optimization of UAV's trajectory is advantageous to the optimization of transmit beamforming vectors in UAV-assisted ISAC systems, which becomes significantly increased for the sufficient mission period, e.g., $T \geq 4$. This is because most of the performance gain obtained by prospoed algorithm results from the design of the UAV's flying path, which becomes pronounced with the sufficiently-large $T$ as shown in Fig. 3. Furthermore, the transmit power constraint cannot be influential in the secrecy performance of \textit{``No transmit beamforming. opt.''} since the higher transmit power makes the stronger data signal, but also the higher information leakage to the eavesdropper as well.
 \vspace{-0.2cm}
\section{Conclusions}
 \vspace{-0.1cm}
In this correspondence, we have developed the UAV-enabled ISAC system with the aim of maximizing the sum secrecy rate, where the UAV is adopted as a base station to serve the multiple ground users, while simultaneously sensing and jamming the passive ground eavesdropper.
Numerical results verify the superiority of the proposed algorithm, which becomes emphasized with the sufficient mission period. As future works, we expand to encompass the more generalized scenarios with multiple UAV and ground eavesdroppers.


%

\ifCLASSOPTIONcaptionsoff
  \newpage
\fi


 \vspace{-0.2cm}
\bibliography{bare_jrnl}
\bibliographystyle{IEEEtran}
%

%




\end{document}